# PULSAR Effect: Revealing Potential Synergies in Combined Radiation Therapy and Immunotherapy via Differential Equations

[1]Samiha Rouf, [2]Casey Moore, [2]Debabrata Saha, [2]Dan Nguyen, [1,3]MaryLena Bleile, [2]Robert Timmerman, [1,2,#]Hao Peng, [1,2,#]Steve Jiang

[1]Medical Artificial Intelligence and Automation Laboratory, University of Texas Southwestern Medical Center, Dallas, TX, 75390, USA

[2]Department of Radiation Oncology, University of Texas Southwestern Medical Center, Dallas, TX, 75390, USA

[3]Department of Statistical Science, Southern Methodist University, Dallas, TX, 75275, USA

E-mail: hao.peng@ utsouthwestern.edu, steve.jiang@utsouthwestern.edu

**Abstract.** PULSAR (personalized ultrafractionated stereotactic adaptive radiotherapy) is a form of radiotherapy method where a patient is given a large dose or "pulse" of radiation a couple of weeks apart rather than daily small doses. The tumor response is then monitored to determine when the subsequent pulse should be given. Pre-clinical trials have shown better tumor response in mice that received immunotherapy along with pulses spaced 10 days apart. However, this was not the case when the pulses were 1 day apart. Therefore, a synergistic effect between immunotherapy and PULSAR is observed when the pulses are spaced out by a certain number of days. In our study, we aimed to develop a mathematical model that can capture the synergistic effect by considering a time-dependent weight function that takes into account the spacing between pulses. By determining feasible parameters, and applying reasonable conditions, we utilize our model to simulate murine trials with varying sequencing of pulses. We successfully demonstrate that our model is simple to implement and can generate tumor volume data that is consistent with the pre-clinical trial data. Our model has the potential to aid in the development of clinical trials of PULSAR therapy.

**Keywords:** PULSAR, immunotherapy, mathematical model, tumor growth, murine trials, tumor microenvironment

# 1      Introduction

For years, conventional radiation therapy has relied on the daily administration of fractionated doses over an extended period. However, recent advancements have unveiled a compelling alternative: strategically spacing the radiation fractions across weeks, ushering in a new era of treatment personalization and adaptability. Studies have indicated a decrease in acute toxicity (Jain et al., 2013) and an improvement in the overall quality of life among specific cancer patient cohorts (Quon et al., 2018) following the adoption of this methodology. This revelation spurred the development of **P**ersonalized, **UL**trafractionated **S**tereotactic **A**daptive **R**adiotherapy (PULSAR).

PULSAR introduces a pioneering method in radiation therapy by employing deliberately spaced large doses, known as 'pulses', of radiation over an extended timeline (Moore et al. 2021). Given its innovative nature, unraveling the essential components contributing to favorable health outcomes stands as a crucial precursor to informed treatment planning and the formulation of clinical trials.

Of particular interest is the intriguing nexus between PULSAR and immunotherapy. Prior research has illuminated the potential immunosuppressive effects of radiation therapy (Sharabi et al., 2015), propelling the fusion of radiation with immunotherapy and demonstrating promising outcomes for select patient cohorts (Zhu et al., 2021). To delve deeper, Moore et al. (2021) conducted pre-clinical trials integrating PULSAR with immunotherapy, revealing a synergistic effect when radiation pulses were strategically spaced.

Within the intricate landscape of the immune system, T-cells stand as pivotal players. Recent investigations have shed light on the dynamic nature of various T-cell populations (e.g., CD8+, Treg cells) in response to radiation exposure (Gough et al., 2022). Notably, studies have revealed shifts in T-cell composition following radiation therapy. Muroyama et al. (2017) reported a notable surge in Treg cells seven days post-radiation, highlighting a proportional increase in this T-cell subtype. Moreover, the work by Arina et al. (2019) observed an intriguing phenomenon wherein intratumoral T-cells, primarily dominated by Treg cells post-irradiation, not only retained their capacity to produce IFN-gamma but exhibited an increase in production spanning 5-9 days post-irradiation. These findings underscore the intricate and multifaceted responses of T-cell subsets to radiation, signaling a compelling avenue for exploring the interplay between radiation modalities and the immune landscape.

One way to begin understanding the complex biological mechanisms that regulate cellular and molecular interactions within the tumor microenvironment (TME) is by employing mathematical models. Typically, these models segment the TME into compartments representing various cells and molecules. For instance, Eftimie et al. (2010) reviewed models focusing on tumor cells and immune effector cells, resembling a predator-prey model (Lotka, 1925). In contrast, Bekker et al. (2022) expanded this framework with a 6-compartment model incorporating antigens and regulatory cells, while Hoffman et al. (2018) explored immune responses concerning antibody and natural killer (NK) cell concentrations. Accounting for the influence of treatment, different approaches have been explored. Watanabe et. al. (2016) proposed a tumor growth model distinguishing active and non-active tumor cells, adjusting growth parameters post-radiation. Serre et al. (2016) employed discrete time equations in tandem with immunotherapy administration.

This paper introduces a mathematical model depicting tumor growth during PULSAR treatment, with or without adjunct immunotherapy. Our model delineates the interaction between tumor cells and T-cells within the TME. Departing from previous models, ours incorporates a time-dependent weight function, capturing two pivotal mechanisms: the induced infiltration of T-cells into the TME and the augmented effector function of existing T-cells. We hypothesize that these mechanisms manifest optimally with a precise interval between radiation pulses. This framework furnishes a simplified yet robust simulation of the PULSAR effect, exhibiting reasonable accuracy. Additionally, we leverage this framework to craft a murine trial simulator, a significant contribution to advancing PULSAR treatment research.

## 2 Mathematical Model

### 2.1 Hypothesis and Assumptions

In seeking to understand the complexities inherent in the tumor microenvironment (TME) during PULSAR treatment with or without immunotherapy, we confront a landscape that remains largely enigmatic. Nevertheless, insights from murine experiments by Moore et al. (2021) suggest a heightened synergistic effect between radiation and immunotherapy, particularly when the pulses are spaced over several days. Building upon this observation, in conjunction with extrapolations drawn from existing literature, our hypothesis posits that radiation potentially triggers the recruitment and activation of T-cells, a process that matures gradually over subsequent days following irradiation. The temporal spacing between pulses appears pivotal, affording sufficient intervals for T-cell recruitment and activation. The intricate web of the body's immune response involves a nuanced interplay among T-cells, antigens, molecular signaling, and regulatory mechanisms (Kim et. al., 2007). To streamline our model, we delineate T-cells into broad categories. Our model is constructed upon the following foundational assumptions.

***Tumor Growth Dynamics:*** The model postulates exponential growth in tumor volume, an observation consistent with empirical tumor volume data in untreated mice. Within our framework, tumor cell demise arises from either irradiation or interactions with T-cells. The removal of deceased tumor cells from the microenvironment occurs at a decaying rate proportional to the tumor volume (Bekker et al., 2022), contributing to the total tumor volume (Watanabe et al., 2016).

***Immune Response and T-cell Dynamics:*** Central to our model lies the representation of the TME's immune response via the interactions between T-cells and tumor cells. We bifurcate T-cells broadly into two categories: newly infiltrated T-cells and intratumoral T-cells (Arina et al., 2019). The former denotes effector cells freshly recruited into the TME, with a subset maturing within the microenvironment to become intratumoral T-cells. Our model assumes a negligible impact of the immune response until the administration of immunotherapy.

***Radiation Influence and T-cell Effector Function:*** We incorporate into our model the premise that irradiation selectively eliminates a fraction of both tumor cells and newly infiltrated T-cells, guided by the LQ model (Barendsen, 1982). In contrast, intratumoral T-cells demonstrate resilience to radiation, while the deliberate spacing of radiation pulses fosters an enhanced effector function in these T-cells.

### 2.2 Model Design

We established a compartmentalized system comprising four populations within the tumor microenvironment (TME): tumor cells $T(t)$, deceased tumor cells $D(t)$, newly-infiltrating T-cells $N(t)$, and intratumoral T-cells $I(t)$. These populations are influenced by two treatment modalities: PULSAR radiation therapy alone (RT) or in conjunction with immunotherapy (RT+PDL1). Figure 1 illustrates the interactions between these cells and the treatments.

The dynamics of each population are governed by a series of coupled ordinary differential equations, as detailed below:

$$\frac{dT}{dt} = \mu T - f_1 TN - f_2 TI \qquad (1)$$

$$\frac{dD}{dt} = f_1 TN + f_2 TI - cD \qquad (2)$$

$$\frac{dN}{dt} = \gamma N - \lambda N - g_1 N + F(t - t_r^i)\rho \qquad (3)$$

$$\frac{dI}{dt} = \lambda N - g_2 I + F\left((t - t_r^i) - \tau\right)\phi \qquad (4)$$

Eqs (1) and (2) describe the growth and mortality of tumor cells in the TME. In Eq (1), the tumor grows at the rate of $\mu$ and killed during interactions with the T-cells at rates $f_1$ and $f_2$. The deceased tumor cells transit to Eq (2) and are cleared from the system at rate $c$.

Eqs (3) and (4) describe the temporal dynamics of two types of T-cells. In Eq (3) newly infiltrated T-cells ($N$) influx the microenvironment at rate $\gamma$. Spacing between pulses of radiation facilitates recruitment by a maximum of $\rho$ cells. The conversion to intratumoral T-cells is represented by $\lambda$. Similarly, intratumoral T-cells ($I$) increase due to the conversion from newly infiltrated T-cells in Eq (4). A maximum of $\phi$ cells are activated once a certain period passes between pulses. The exhaustion rates of $N$ and $I$ are $g_1$ and $g_2$, respectively. $F$ is a time-dependent weight function that accounts for the tie between pulses (Refer to Section 2.3). it controls the level of recruitment of $N$ in Eq (3), and the level of improved effector function of $I$ in Equation (4). $t_r$ represents the series of days that the pulses of radiation were administered, where $i$ represents the index of the pulses. For instance, if there were two pulses, one given on day 14 and the next given ten days later then $i \in \{1,2\}$ and $t_r = \{t_r^1, t_r^2\} = \{14, 24\}$. $\tau$ is a fixed time delay in the range between 5 and 9 days post irradiation (Arina et. al. 2019).

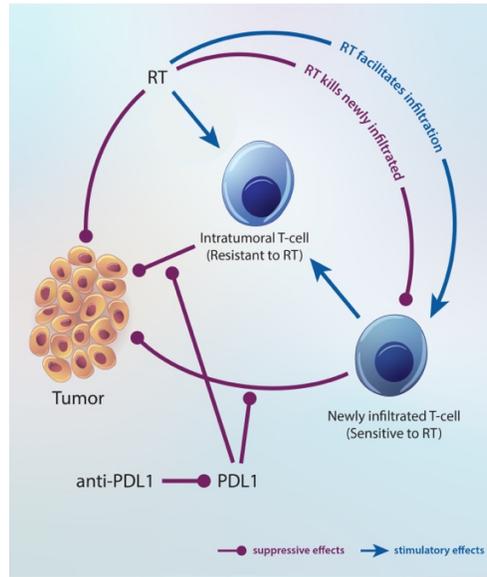

**Fig. 1.** Visualization detailing tumor-tissue dynamics involving tumor cells, T-cells, radiation (RT), and anti-PDL1 immunotherapy in the microenvironment.

**The RT-only model:** Since there is no immunotherapy, we assume the immune response is negligible. Eqs (1)-(4) reduces to Eqs (5)-(6).

$$\frac{dT}{dt} = \mu T \qquad (5)$$

$$\frac{dD}{dt} = -cD \qquad (6)$$

**The RT+PDL1 model:** For the treatment plans involving PULSAR and immunotherapy the model returns to the original equations described by Eqs (1)-(4). The immunotherapy, α-PD-L1, is first administered two days before each pulse and then every two days post-irradiation. Since the first dose of α-PD-L1 occurs before irradiation, then initially the time-dependent weight function $F = 0$ in Eqs (3)-(4). After irradiation, function $F$ is triggered as described in Section 2.3. Survival fraction after irradiation is commonly determined with the LQ model $S(d) = e^{-\alpha d - \beta d^2}$ (Barendsen 1982), where $d$ is the dose of radiation, and $\frac{\alpha}{\beta}$ is the measure of intrinsic radiosensitivity of a tissue. Similar to the framework utilized by Watanabe et. al. (2016), once a pulse of radiation is administered, the tumor cells $T(t)$ and newly-infiltrating T-cells $N(t)$ at time $t^-$ drops to its surviving portion at time $t^+$. The remaining fraction of tumor cells is presumed dead, contributing to the volume of deceased tumor cells $D(t)$. Therefore, from time $t^-$ to time $t^+$ we have:

$$T(t^+) = S(d) \cdot T(t^-) \qquad (7)$$

$$D(t^+) = (1 - S(d)) \cdot T(t^-) + D(t^-) \qquad (8)$$

$$N(t^+) = S(d) \cdot N(t^-) \qquad (9)$$

Numerical implementation of the LQ model is further described in Section 3.3 and illustrated in Figure 2.

## 2.3   The weight function *F(t)*

We propose that a significant interval between radiation pulses leads to an augmented recruitment of newly infiltrated T-cells and an enhanced effector function of intratumoral T-cells. Consequently, a time-dependent response is triggered by radiation exposure. To mathematically capture this temporal response, we introduce the weight function F(t) into Eqs (3) and (4). Employing the hyperbolic tangent function F(t) = tanh(t) in our model signifies that, for t≥0, F(t) steadily increases from 0 to 1 over time.

In Eq (3), ρ defines the maximal rate of recruitment for newly infiltrated T-cells. Consequently, when the interval between pulses is brief, F(t) remains small, resulting in a reduced cell recruitment rate. Conversely, with a longer interval, F(t) approaches 1, fostering a higher recruitment rate, up to the maximum set by ρ. Turning to Eq (4), the effector function of intratumoral T-cells undergoes augmentation 5-9 days post-irradiation. We model this enhanced function by considering a subset of dormant intratumoral T-cells transitioning into an activated state within the TME. Thus, ϕ delineates the maximum activation rate achievable by intratumoral T-cells between radiation pulses. This maximum activation is attained when τ exceeds the interval between pulses (F(t) nears 1). Alternatively, if τ is shorter than the interval between pulses, there's no escalation in the effector function level.

## 3   Methods and Experiments

### 3.1   Murine study design and data

In this study, female C57BL/6J mice, aged six to eight weeks, were specifically chosen for their compatibility with the Lewis lung carcinoma (LLC) cell lines derived from lung cancer in C57BL/6 mice. To induce tumor growth, $1 \times 10^6$ LLC cells were subcutaneously injected into the right legs of the mice. Monitoring commenced until tumors reached a standardized size range of 150 to 200 $mm^3$, at which point the mice were randomly assigned to distinct treatment groups.

The treatment groups were administered either α-PD-L1 or the isotype control, following randomization upon reaching the predetermined tumor size. Tumor progression was assessed by measuring length, width, and height using a caliper, with mice reaching humane endpoints upon exhibiting tumor volumes surpassing 1500 $mm^3$ or displaying significant tumor ulceration, necessitating euthanasia.

Conducting the experiments in three iterative rounds allowed for a comprehensive collection of data. Leveraging a subset of the data previously established by Moore et al. (2021), our focus lay in selecting specific treatment plans for model fitting and validation. We specifically targeted treatment plans involving two pulses of 10 Gy, as per the scheme denoted as 10GyD# or 10GyD#+PDL1, where # represents the interval in days between administered pulses and +PDL1 signifies concurrent immunotherapy application.

For instance, the designation 10GyD4+PDL1 corresponds to experiments where mice received dual 10 Gy pulses separated by a 4-day interval alongside immunotherapy. Refer to Table 1 for a comprehensive catalog detailing the treatment plans meticulously employed for model calibration and validation.

**Table 1.** List of treatment plans with two pulses.

| RT Dose | Round 1 | Round 2 | Round 3 |
|---|---|---|---|
| No RT (0Gy) | Vehicle | Vehicle | Vehicle |
|  | PDL1 only | PDL1 only | PDL1 only |

| 10Gy | D1 | D1 | D4 |
| --- | --- | --- | --- |
|  | D1+PDL1 | D1+PDL1 | D4+PDL1 |
|  | D10 | D-10 |  |
|  | D10+PDL1 | D10+PDL1 |  |

*D{#} indicates the days between pulses*
*+PDL1 indicates that immunotherapy was given*

Each treatment plan outlined in Table 1 was administered to a cohort consisting of 5-9 mice. To ensure a comparative analysis across the various treatment plans, we calculated the average tumor volumes within each cohort. These averaged tumor volumes formed the datasets employed for both the fitting and validation stages of our model.

### 3.2 Numerical implementation

To solve the mathematical models, namely the RT model and RT+PDL1 model, we employed the ode45 solver within MATLAB (The MathWorks Inc., Natick, MA). Our utilization of event functions within the solver facilitated halting computations at each radiation pulse and during the onset or conclusion of immunotherapy effects.

Our approach involved the direct integration of tumor volume data into our model equations. To achieve this, we calibrated the equations governing tumor cells (Eqs. (1) and (2)) and their associated parameters by adopting a scaling premise wherein 1 cubic millimeter of tumor volume approximately equates to 100 thousand cells (Concurrent Series 2012). This adjustment necessitated reformulating the solutions for Eqs. (1)-(2) as follows: T(t) denoting tumor volume in $mm^3$ and D(t) representing the volume of deceased tumor cells in $mm^3$. Additionally, for ease of computation, equations and parameters governing T-cells (Equations (3) and (4)) were scaled by a factor of $10^3$ cells. During the fitting process, the selected initial conditions are $\{T_0, D_0, N_0, I_0\} = \{5, 0, 0.1, 0.1\}$. After fitting the model, each $T_0$ was fine-tuned iteratively to minimize the root mean square error (RMSE). Total tumor volume on any given day $t$ is computed by taking the sum of $T(t)$ and $D(t)$.

An illustrative understanding of the relationship between T(t) and D(t) can be conceived by considering a tumor with a survival fraction of 25% after ten days, i.e., $T(10^-) = 250\ mm^3$, with no deceased tumor cells initially, i.e., $D(10^-) = 0$. Post a radiation pulse on Day 10, resulting in 187.5 $mm^3$ of tumor cell death, we observe $D(10^+) = 187.5\ mm^3$ and $T(10^+) = 62.5\ mm^3$. Figures 3, 4, 5, and 6 visually encapsulate the total tumor volume dynamics.

The implementation of the LQ model was contingent upon scheduling the pulse days as stopping criteria for MATLAB's event functions. Upon reaching these predefined time events, the solver ceased computations and recalculated new initial conditions utilizing Eqs (7)-(9). Specifically, in treatment plans involving both immunotherapy and radiation, the system of equations transitioned from (5)-(6) to (1)-(4) upon α-PD-L1 administration. This transition was scheduled two days before the first radiation pulse. After computing new initial conditions using the LQ model, the solver resumed solving Eqs (1)-(4) for the remaining simulation duration. Figure 2 depicts this workflow of the solver.

Considering the assumption of intratumoral T-cells' radio-resistance, the LQ model was not applied to I(t). Moreover, maintaining consistency, the same α and β values were applied to both tumor cells and newly infiltrated T-cells.

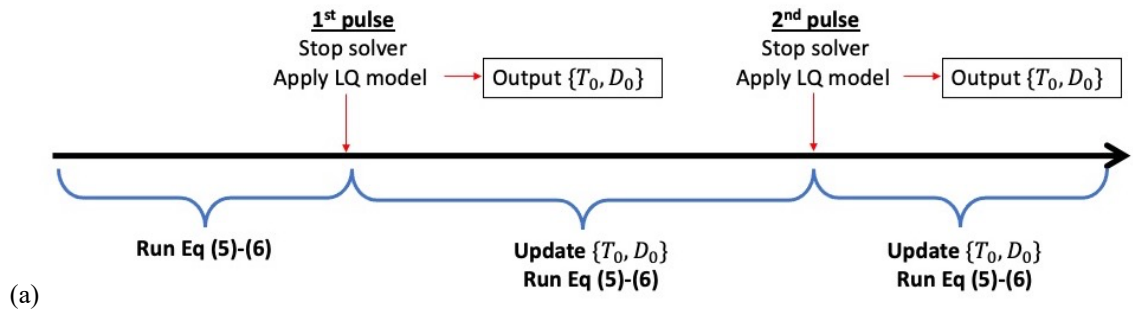

(a)

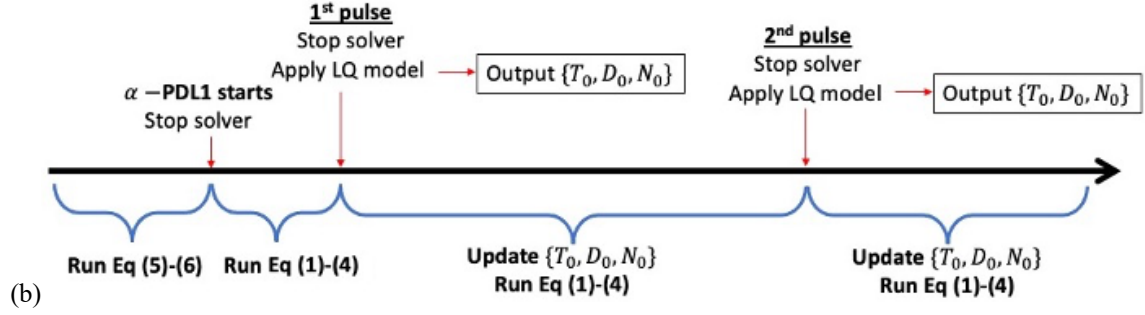

(b)

**Fig. 2.** Visually depicting solver workflow and interruptions (a) RT-only treatment plans (b) RT with α-PDL1 treatment plans.

### 3.3 Parameter fitting

We selected the treatment plans from Round 2 to fit the model parameters. Parameters were fitted using a simulated annealing algorithm (Press et. al., 2007) to minimize the cumulative sum of the normalized mean-square errors (MSE) of the selected treatment plans. We fit the parameters in three stages. In each stage, we select $n$ treatment plans and the related set of parameters, $\theta$. Each plan has a set of averaged measured tumor volumes with which we computed the MSE in relation to the predicted tumor volume at the corresponding time step. The MSE is normalized by the length of the corresponding treatment period, $\{p_1, p_2 \dots p_n\}$. Then we find the sum of the normalized MSE for the selected treatment plans. Therefore, we have the following minimization problem:

$$\min_{\theta \in \Theta} \sum_{i=1}^{n} \frac{MSE_i}{p_i}$$

Θ represents the reasonable parameter values from which values are selected for the parameter set $\theta$.

***Stage 1*** Fitting tumor growth parameter $\mu$. When no treatment is applied (Refer to Equation 5) the only parameter at play is $\mu$ which represents the tumor growth rate. Hence, we determined this rate by fitting the Vehicle data of Round 2. We fix the value for $\mu$ for the rest of the experiments.

***Stage 2*** Fitting RT model parameters $\theta = \{\alpha, \beta, c\}$ to treatment plans {10GyD1, 10GyD10}. Since $\mu$ was determined in Stage 1 we are now concerned with fitting the LQ model parameters of $\alpha$ and $\beta$ and the deceased tumor cells removal rate $c$. We set the condition for the ratio $\alpha/\beta \leq 10$ (Watanabe et. al. 2016, van Leeuwen 2018). These parameter values are then fixed.

***Stage 3*** Fitting RT+PDL1 model parameters that remain to treatment plans {PDL1 only, 10GyD1+PDL1, 10GyD10+PDL1}. These parameters govern the mechanisms and interactions of the newly infiltrating and intratumoral T-cells. We also fit the delay time parameter $\tau$ such that $5 \leq \tau \leq 9$. Therefore, we have $\theta = \{\gamma, \lambda, \rho, \phi, f_1, f_2, g_1, g_2, \tau\}$.

The fitted parameters (Refer to Table 2) were fixed and then applied to the treatment plans of Round 1 and Round 3. These same parameters were used for the mice trial simulator.

### 3.4 Fine-tuning initial tumor volume

As tumor volume was assessed using a caliper post-implantation, the initial tumor volume $T_0$ remained unmeasurable and was consequently recorded as $0\ mm^3$ for each mouse in the study. Nonetheless, to enable the application of the ode solver within the model, a nonzero initial volume was imperative. Hence, during model fitting, we standardized $T_0 = 5\ mm^3$ for every treatment plan. Realistically, the initial volume would exhibit variability among individual mice. Subsequent to parameter fitting, an iterative process ranging $T_0$ between $1\ mm^3$ to $10\ mm^3$ was undertaken to ascertain the initial volume that minimized the root mean square error (RMSE) for each treatment plan. Figure 3 outlines the outcomes from fine-tuning $T_0$ for the treatment plan 10GyD10. Notably, the model initially set with an initial tumor volume of 5 $mm^3$ exhibited an RMSE of 117.91, which significantly reduced to 34.59 following the fine-tuning of $T_0$.

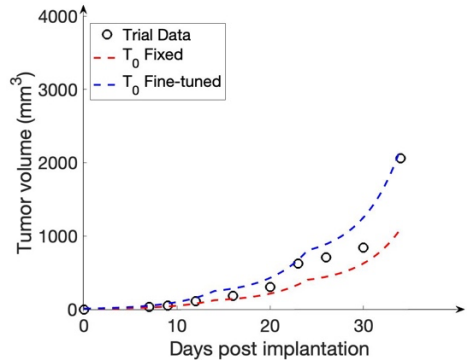

**Fig. 3.** Average volumes of mice in the 10GyD10 trial are indicated by black circles. The model initially set with $T_0$ as 5 $mm^3$ (dotted red line) yielded an RMSE of 117.91, while the model with fine-tuned initial volume of 8.6 $mm^3$ (dotted blue line) yielded an RMSE of 34.59.

### 3.5 Designing Trial Simulator

Our model was instrumental in crafting a simulator designed to generate tumor volume outputs for mice subjected to either PULSAR treatment or PULSAR combined with immunotherapy. Introducing variability among simulations, the tumor growth rate (μ) was randomly selected within the range of 0.2 to 0.3. Additionally, initial tumor volumes were diversified, spanning between 1 $mm^3$ to 10 $mm^3$, while all other parameters remained fixed as outlined in Table 2. Leveraging our model, we projected the tumor volume trajectories for mice across a 40-day period in these trials, adhering to a humane endpoint wherein mice were euthanized upon reaching a volume surpassing 1500 $mm^3$, even if the 40-day duration had not transpired.

Our focus was directed towards simulating trials with varied intervals between the two administered pulses. Initiating the first pulse on day 14, each pulse maintained a fixed dose of 10 Gy throughout every simulation. Subsequently, the simulator generated paired tumor volume data for each treatment type (RT only and RT+PDL1) across diverse pulse spacings. Specifically, we identified a 'PULSAR effect' when the final day's tumor volume of mice receiving RT only surpassed that of mice undergoing RT+PDL1 treatment on the same day. This difference was quantified and recorded for every simulator run. Figure 4 elucidates the outcomes from a singular simulator run where pulses were spaced by 9 days which yielded a PULSAR effect of magnitude 590 $mm^3$ on the 37[th] day.

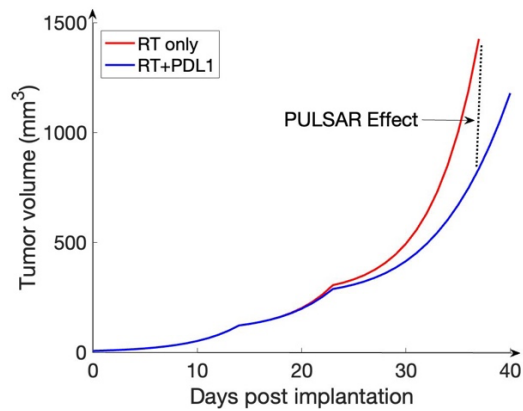

**Fig. 4.** With a pulse spacing set at 9 days, the simulator presents tumor volume data for PULSAR-only (solid red line) and PULSAR with immunotherapy (solid blue line) treatment plans. In this simulation μ=0.22 and $T_0 = 5.7$.

## 4 Results

### 4.1 Fitting Parameters

As described in Section 3.4, the treatment plans of Round 2 were fitted in three stages via a simulated annealing algorithm. Table 2 lists the best-fit values obtained for each parameter.

Table 2. Parameter values

| Parameter | Value | Unit | Description |
|---|---|---|---|
| $\mu$ | 0.23 | $mm^3 day^{-1}$ | Growth rate of T |
| $\alpha$ | 0.0229 | $Gy^{-1}$ | Radiosensitivity factor in the LQ model |
| $\beta$ | 0.0114 | $Gy^{-2}$ | Radiosensitivity factor in the LQ model |
| $c$ | 0.0087 | $mm^3 day^{-1}$ | Clearance rate of D |
| $f_1$ | 0.0110 | $C^{-1} day^{-1}$ | Cell death by newly recruited T cells (N) |
| $f_2$ | 0.0107 | $C^{-1} day^{-1}$ | Cell death by intratumoral T cells (I) |
| $\gamma$ | 0.0097 | $C \cdot day^{-1}$ | Influx rate of N |
| $\lambda$ | 0.0107 | $C \cdot day^{-1}$ | Conversion rate from N to I |
| $\rho$ | 1.05 | $C$ | Recruitment of N induced by radiation |
| $\phi$ | 1.02 | $C$ | Factor due to enhanced effector function |
| $g_1$ | 0.5613 | $C \cdot day^{-1}$ | Exhaustion rate of N |
| $g_2$ | 0.1000 | $C \cdot day^{-1}$ | Exhaustion rate of I |
| $\tau$ | 5 | $days$ | Delay window for the enhanced effector function |

$C$: $10^3 \, cells$

T - tumor volume, D - deceased tumor cells, N - newly infiltrating T-cells, I - intratumoral T-cells

We fit the parameters, α and β, recognizing that survival curves are not well described by the LQ model for large dose ranges (Park et al., 2008). Numerically determining these values became pertinent. Moreover, during the fitting process, we adhere to the condition that the α/β ratio should not surpass the common threshold of 10 for tumors (Watanabe, 2016; van Leeuwen et al., 2018). Consequently, our fitting process yielded an α/β ratio of approximately 2, indicating a substantial damaging effect of radiation on the tumor.

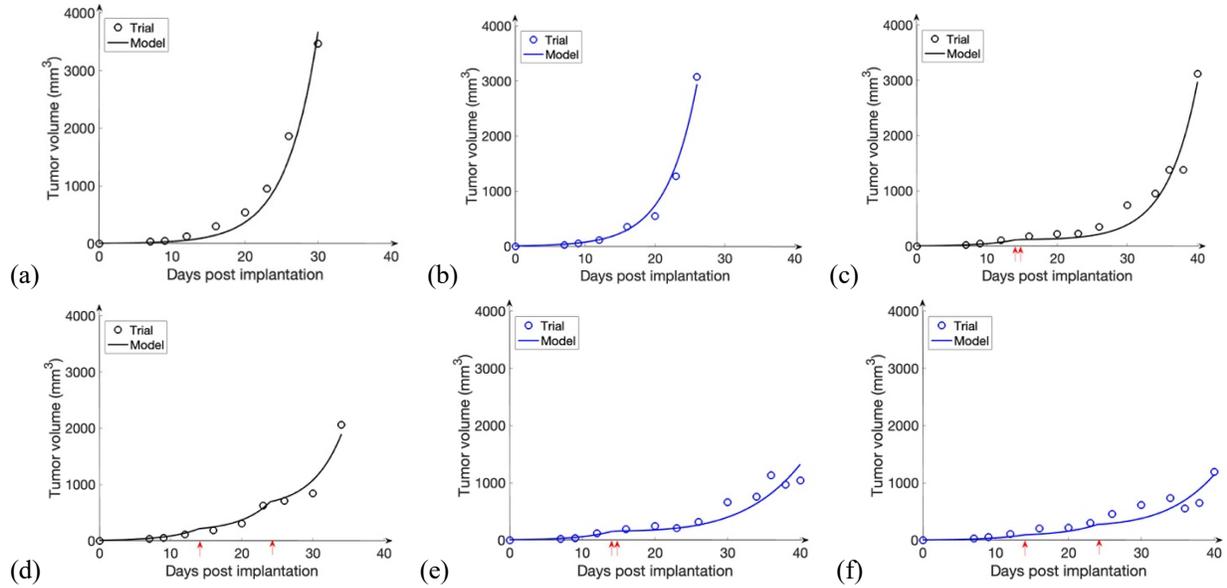

**Fig. 5.** The circles represent the average of the measured tumor volume from the murine trials of Round 2. The solid lines represent the tumor volume trajectory of the model after determining the best-fit parameters. Red arrows indicate the day a pulse of RT was given. Blue plots indicate that immunotherapy was administered. **(a)** Vehicle with RMSE 68.37 **(b)** PDL1 only with RMSE 45.98 **(c)** 10GyD1 with RMSE 58.93 **(d)** 10GyD10 with RMSE 34.60 **(e)** 10GyD1+PDL1 with RMSE 43.97 **(f)** 10GyD10+PDL1 with RMSE 38.36.

In Figure 5, the tumor volume trajectories depicted represent the cumulative sum of T(t) and D(t) outputs from our model. Figures 5(a) and (b) illustrate that in the absence of radiation, tumor growth remains exponential, even if α-PDL1 is present. However, upon comparing Figures 5(c) to (e) and 5(d) to (f), the impact of immunotherapy becomes notably pronounced following radiation administration. The model's trajectories closely mimic the observed data trends, demonstrating reasonable accuracy—highlighted by an average RMSE of 48.37 across the six plots. Furthermore, the externally measured tumor volume exhibits no substantial drops after irradiation, as seen in Figures 5(c), (d), (e), and (f). Our model accounts for this behavior by incorporating a fitted low clearance rate of deceased tumor cells, indicating a gradual decline in the total tumor volume post-irradiation.

### 4.2   Model Validation

To assess the model's performance against murine trial data, we utilized the parameter values delineated in Table 2, employing them to simulate the treatment plans featured in Round 1 and Round 3. The resultant model outputs for each treatment plan are depicted in Figure 6. Noteworthy is the Root Mean Square Error (RMSE) range, spanning from 20 to 100 and averaging at 65.7.

Observing specific experiments, discernible disparities between the data and model output become apparent, prominently highlighted in Figure 6 (c, d, e, j), where the RMSE surpassed 75. These discrepancies likely stem from errors introduced during data imputation, necessary for estimating tumor volumes in instances where mice either perished or were sacrificed before trial completion.

Particularly intriguing is the consistent tumor volume increase observed for treatment plans 10GyD1 and 10GyD1+PDL1, even post-back-to-back radiation pulses, followed by near-constant growth (refer to Figure 6 c, d). Intriguingly, the model's trajectory fails to mirror this growth pattern for these plans. However, when the pulses were spaced 4 days (Figure 6 i, j) or 10 days apart (Figure 6 e, f), the model's trajectory more closely aligned with measured data.

This discrepancy might be attributed to the inherent limitations of the LQ model in determining the survival fraction for larger doses. The close succession of pulses, mimicking a single large dose, might lead to an overestimation of the survival fraction, consequently impacting the model's ability to capture this specific growth pattern.

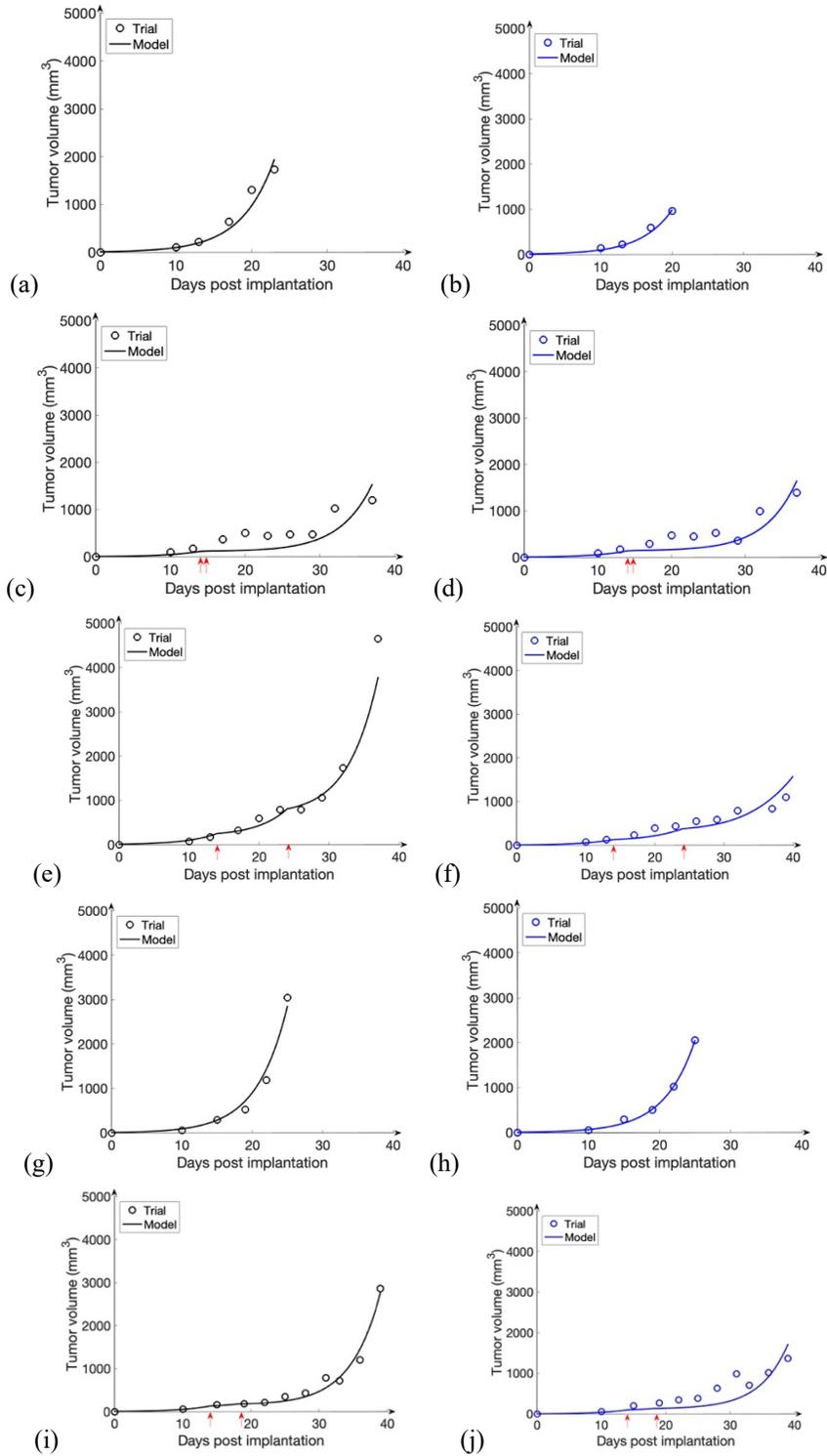

**Fig. 6.** Plots **(a)-(f)** are the results for Round 1: Vehicle, PDL1 only, 10GyD1, 10GyD1+PDL1, 10GyD10, 10GyD10+PDL1 respectively. Plots **(g)-(j)** are the results for Round 3: Vehicle, PDL1 only, 10GyD4, 10GyD4+PDL1 respectively. The circles represent the average of the measured tumor volume from the murine trials. The solid lines represent the tumor volume trajectory of the model. Red arrows indicate the day a pulse of RT was given. Blue plots indicate that immunotherapy was administered.

### 4.3 Trial simulator

The mice trial simulator conducted 100 simulations, varying pulse spacings from 1 to 20 days for two treatment plans: PULSAR alone and PULSAR combined with immunotherapy. This generated a total of 4000 simulated trials. Simulation cessation criteria were set at a tumor volume of 1500 $mm^3$ or the elapse of 40 days, determining the final day of each trial based on whichever condition occurred first.

Subsequently, the disparity in tumor volume on the final day between treatment pairs was computed across the 100 simulations and averaged. These averages are visualized in Figure 7. The primary aim was to gauge the PULSAR effect's magnitude concerning pulse spacing. A larger magnitude denotes a more pronounced reduction in tumor volume over the same duration, signifying a favorable health outcome.

Figure 7 demonstrates a discernible trend: as pulse spacing increases from 1 to 8 days, a gradual rise in the average difference's magnitude is observable. Notably, once the spacing extends to 9 days, a significant surge in magnitude is evident. Subsequently, the magnitude fluctuates periodically between 200 $mm^3$ to 250 $mm^3$ as the spacing further increases.

These outcomes suggest that spacing pulses by 9 days or more may induce a PULSAR effect. This aligns with Moore et al.'s findings (2021), where they observed a similar effect when pulses were spaced 10 days apart.

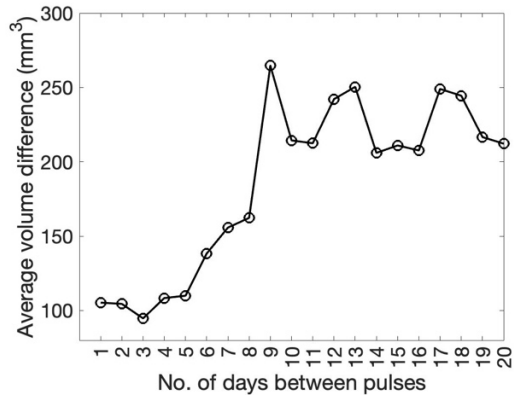

**Fig. 7.** The plot illustrates the average PULSAR effect magnitude at varying pulse spacings. This effect measures the difference in tumor volume between trials receiving PULSAR alone and those combined with immunotherapy at trial completion. Notably, the plot highlights that pulse spacings of 9 days or more exhibit a substantial PULSAR effect.

## 5 Discussion

The results of pre-clinical experiments have revealed the synergy between immunotherapy and PULSAR treatment when the administration of pulses is spaced over a defined timeframe. However, the underlying mechanisms driving this synergy remain unclear. To unravel the intricate biological processes at play, we developed a mathematical model that comprehensively considers the dynamic interplay between tumor cells and T-cells within the tumor microenvironment (TME). The model is capable of generating tumor volume trajectories under various sequences of PULSAR treatment, with or without immunotherapy.

As evidenced by the results presented in Figure 6, the predicted temporal change of tumor volume align closely with the experimental data, reflecting the model's accuracy. Furthermore, the mathematical framework can be used as an in-silico simulator, offering a unique opportunity for investigating potential PULSAR effect for new combinations.

Figure 7 illustrates a notable outcome, suggesting that the synergistic effect between PULSAR treatment and immunotherapy is most pronounced when pulses are spaced out by 9 days or more.

One major advantage of our proposed framework is its inherent flexibility. It allows for the manipulation of various treatment parameters, including altering the sequencing of pulses, adjusting the timing of immunotherapy administration, and modulating radiation dosage. Moreover, its simplicity in implementation, utilizing well-established ordinary differential equation solvers, adds to its practicality. A number of fitted parameters, particularly those related to T cell migration and infiltration, can be validated and fine-tuned through experiments including flow cytometry and molecular imaging technologies. For example, optical imaging tools can be employed to investigate the dynamics of T cells, comprising their migration from lymph nodes to the targeted tumor via the circulatory system and their subsequent infiltration into the tumor. By incorporating additional compartments to account for the diverse cells and molecules within the immune system, our model can evolve into a more comprehensive representation of the immune response to various treatments. Another advantage of our proposed model over AI-driven approaches lies in their interpretability, making them valuable tools for augmenting the capabilities of AI frameworks (Benzekry, 2020). For example, the mathematical framework we developed may seamlessly complement AI models by providing simulated input data. In the long term, one potential research avenue is the development of digital twins (Sun et. al., 2022; Erol et. al., 2020). Digital twins are virtual environments designed to replicate physical objects, encompassing a multitude of processes that simulate in real-time using data-driven insights and machine learning, spanning the entire lifecycle of the system.

Several limitations, arising from the preliminary results outlined in this manuscript, require further attention. Firstly, increasing the number of animals in each group could diminish intra-group variation and yield a more accurate estimation of mean tumor volume, especially at the end of the experiments. Secondly, extending the same framework to assess different cell lines beyond LLC is essential for generalization. Diverse behaviors are expected due to variations in tumor microenvironment and radiosensitivity. Thirdly, the measurement of only tumor volume change is a limitation. In other words, the definition of synergy and PULSAR effect rely solely on tumor size in our current study, lacking biological correlates such as experimental data on immune cell infiltration and their activities.

Fourthly, a critical assumption in our model pertains to the radiosensitivity of intratumoral T-cells, known to exhibit greater radioresistance compared to newly infiltrated T-cells (Arina et al., 2019). We chose not to employ the LQ model for this particular subgroup of T-cells and, instead, assumed complete radioresistance. While we did examine a few cases in our study by incorporating the LQ model, our findings indicated that the killing of intratumoral T-cells through irradiation has a minimal impact on tumor volume. Additional validation is required to affirm the accuracy of this assumption.

## 6 Conclusion

PULSAR therapy represents a significant step towards personalized medicine, especially when combined with immunotherapy. We have successfully captured the elusive synergistic effect observed in pre-clinical data by developing a mathematical model. Our innovative approach focused on the dynamics of T cells and incorporated time-dependent weight functions. Embracing this unique strategy not only unveiled the intricacies of the PULSAR effect but also demonstrated the model's utility in simulating mice trials with diverse sequencing of radiation pulses. Our model may become an indispensable guide in designing smart trials involving PULSAR more cost-effectively, contributing to the optimization of treatment outcomes.